# Landau Quantization in Graphene Monolayer, Bernal Bilayer, and Bernal Trilayer on Graphite Surface

Long-Jing Yin[§], Si-Yu Li[§], Jia-Bin Qiao, Jia-Cai Nie, Lin He[*]

Electronic properties of surface areas decoupled from graphite are studied using scanning tunnelling microscopy and spectroscopy. We show that it is possible to identify decoupled graphene monolayer, Bernal bilayer, and Bernal trilayer on graphite surface according to their tunnelling spectra in high magnetic field. The decoupled monolayer and bilayer exhibit Landau quantization of massless and massive Dirac fermions, respectively. The substrate generates a sizable band gap, ~35 meV, in the Bernal bilayer, therefore, the eightfold degenerate Landau level at the charge neutrality point is split into two valley-polarized quartets polarized on each layer. In the decoupled Bernal trilayer, we find that both massless and massive Dirac fermions coexist and its low-energy band structure can be described quite well by taking into account only the nearest-neighbor intra- and interlayer hopping parameters. A strong correlation between the Fermi velocity of the massless Dirac fermions and the effective mass of the massive Dirac fermions is observed in the trilayer. Our result demonstrates that the surface of graphite provides a natural ideal platform to probe the electronic spectra of graphene layers.

Department of Physics, Beijing Normal University, Beijing, 100875, People's Republic of China.
[§]These authors contributed equally to this paper.
[*]helin@bnu.edu.cn

The surface monolayer/few-layer graphene may decouple from graphite (or graphene multilayer), which consists of stacked layer of graphene sheets held together by weak van der Waals forces. Therefore, it is possible to detect two-dimensional Dirac fermions on the surface of such three-dimensional systems [1-7]. The richness of possible stacking configurations and the layer number degree of freedom of the decoupled area enable that the surface of graphite (or graphene multilayers) exhibits remarkably complex phenomena and unusual properties [1-3,8-12]. For example, coexistence of massless and massive Dirac fermions has been frequently observed at highly oriented pyrolytic graphite (HOPG) surface [1,3,5,7,13]. However, a systematic experimental study in establishing the close connections between the structures and the electronic properties of the graphite surface is still missing. Lacking of efficient method to simultaneous identify the stacking order, the layer number, and the electronic properties of the decoupled area limits further efforts in exploring the nature of quasiparticles on the surface of graphite [3].

In this Letter, we employ scanning tunnelling microscopy (STM) and spectroscopy (STS) to investigate the surface of highly oriented pyrolytic graphite (HOPG). We find that decoupled graphene monolayer, Bernal bilayer, and Bernal trilayer on HOPG surface can be identified by their tunnelling spectra in high magnetic field. Landau quantizations of massless and massive Dirac fermions are observed in the decoupled monolayer and bilayer, respectively. The substrate (HOPG) breaks the layer symmetry of the Bernal bilayer and generates a sizable band gap, ~35 meV, in this system, which results in valley polarization of the lowest Landau levels on each layer. In the decoupled Bernal trilayer, the tunnelling spectra in high magnetic field consist of Landau quantizations of both massless and massive Dirac fermions. Surprisingly, a strong correlation between the Fermi velocity of the massless Dirac fermions and the effective mass of the massive Dirac fermions is observed in the trilayer.

The HOPG was surface cleaved with adhesive tape prior to experiments. Figure 1 summarizes two typical STM images and 4 typical STS spectra recorded in high magnetic field (8 T) on HOPG surface. Figure 1(a) and 1(b) show the two different topographies with atomic-resolution images of the honeycomb lattice and triangular lattice, respectively. The appearance of the honeycomb lattice on HOPG surface can be regarded as a signature that the top layer is a graphene monolayer decoupled from the bulk HOPG (the separation between the top and second layer is larger than the equilibrium distance 0.34 nm, see Supplementary Fig. S1 for a typical STM image) [4]. The high magnetic field STS recorded in such a region, as shown in the top panel of Fig. 1c, really exhibits non-equally-spaced energy-level spectrum of Landau levels (LLs), which are unique to massless Dirac fermions in graphene monolayer [4,6,9,14].

The triangular contrast in Fig. 1(b) suggests that the top layer of this region is, at least, coupled to the second graphene layer [4]. Usually, the main part of HOPG is the Bernal (AB-stacked) graphite and the A/B atoms' asymmetry generated by the adjacent AB-stacked two layers results in the triangular lattice. Therefore, the triangular lattice is observed more frequently than the honeycomb lattice on HOPG surface. The bright spots of the triangular lattice, as shown in Fig. 1(b), are the sites in the adjacent two layers where one sublattice of the first layer lies above the center of the hexagons in the second layer. Here we should point out that graphene bilayer and trilayer with stacking fault [8-11,15,16], i.e., the twisted bilayer and trilayer, on graphite surface are not included in this work. The twisted bilayer and trilayer exhibit quite different properties comparing with that of the Bernal allotropes and they can be identified easily in the STM images due to the emergence of moiré pattern [8-11,15,16]. Three different high magnetic field STS spectra, as shown in Fig. 1(c), are observed in the regions showing triangular contrast. One of them shows no discernible peak [the bottom panel of Fig. 1(c)], attributing to the spectrum of bulk graphite, i.e., the surface graphene few-layer is coupled strongly to the graphite and it does not exhibit any characteristic of two-dimensional systems. The other two different spectra show pronounced peaks of discrete LLs, indicating the quasi-two-dimensional nature of these surface regions where the spectra are recorded. We attribute the peaks of the two spectra to the Landau quantizations in the decoupled Bernal bilayer and trilayer respectively, the reason of which will be discussed subsequently.

To further study the Landau quantization of graphite surface, we carried out STS measurements on different decoupled surface regions with different magnetic fields. Figure 2(a) shows a series of tunnelling spectra for various magnetic fields from 0 to 8 T recorded on a decoupled graphene monolayer on HOPG surface (the corresponding atomic-resolution STM image shows the honeycomb lattice, see Supplementary Fig. S1). The LLs developed as the magnetic field is increased and the Landau-level peaks are well resolved up to $n = 5$ in both electron and hole sectors (here $n$ is the LL index). Our observation of the LLs is highly similar to that observed on a graphene flake on graphite [4]. For massless Dirac fermions in graphene monolayer, the LL energies $E_n$ depend on the square-root of both level index $n$ and magnetic field $B$ [4,6,9,14]

$$E_n = \text{sgn}(n)\sqrt{2e\hbar v_F^2 |n| B} + E_0, \qquad n = ...-2,\ -1,\ 0,\ 1,\ 2... \qquad (1)$$

Here $E_0$ is the energy of Dirac point, $e$ is the electron charge, $\hbar$ is the Planck's constant, and $v_F$ is the Fermi velocity. Our analysis, as shown in Fig. 1(b) and Fig. 1(c), demonstrates that the sequence of the observed LLs can be described quite well by Eq. (1). The linear fit of the experimental data to Eq. (1) yields a Fermi velocity of $v_F = (1.207 \pm 0.002) \times 10^6$ m/s. The observation of a single sequence of LLs of the massless Dirac fermions demonstrates the efficient decoupling of the surface graphene monolayer from the graphite [4].

Figure 3(a) shows a sequence of d$I$/d$V$ spectra with various magnetic fields obtained in a decoupled region showing the triangular contrast (see Supplementary Fig. S2 for the STM image). In zero magnetic field, there is a pronounced peak located at about 50 mV, which is attributed to the density of states (DOS) peak generated at the conduction band edge of Bernal bilayer [11,17]. With increasing magnetic field, the spectra develop into a sequence of well-defined LL peaks and the DOS peak at the conduction band edge becomes a valley-polarized quartet, $LL_{(0,1,+)}$, mainly localized on the topmost graphene layer (here 0 and 1 are the LL indices and the symbols +/- are valley indices) [11,16-18]. The other valley-polarized quartet $LL_{(0,1,-)}$, residing on the second layer, is much weaker in the tunnelling spectra, as the STM probes predominantly the top layer.

To identify the orbital index $n$ of each LL, we further examine the magnetic field dependence of the LL peak positions, as shown in Fig. 3(c). Besides the two valley-polarized quartets (which are almost independent of magnetic field), the other LLs show linear field dependence. For massive Dirac fermions in Bernal graphene bilayer, the LL spectrum takes the form [3,16,17,19]:

$$E_n = \pm \left[ \hbar \omega_c (n(n-1))^{1/2} + E_g / 2 \right], \qquad n = 0, 1, 2... \qquad (2)$$

Here $\omega_c = eB/m^*$ is the cyclotron frequency, $m^*$ is the effective mass of quasiparticles, and $E_g$ is the band gap. The Landau quantization shown in Fig. 3 provides compelling evidence that the decoupled surface region is Bernal graphene bilayer. A single sequence of LL peaks corresponding to massive Dirac fermions on graphite surface, although anticipated to be observed for many years, had never been experimentally observed before. A fit of experimental data to Eq. (2), as shown in Fig. 3(b), yields the effective mass $(0.041 \pm 0.001)m_e$ for electrons, $(0.046 \pm 0.001)m_e$ for holes (here $m_e$ is the free-electron mass), and the band gap $E_g = 35$ meV. Both the effective mass of the massive Dirac fermions and the band gap are in good agreement with the range of values reported previously for Bernal graphene bilayer on various substrates [3,16,17,20-22].

Besides the high field spectra exhibiting the single sequence of LL peaks of either massless or massive Dirac fermions, we frequently observed another series of spectra, which show sequence of LL peaks of both massless and massive Dirac fermions, on graphite surface. Figure 4(a) shows a series of these d$I$/d$V$ spectra with various magnetic fields (see Supplementary Fig. S3 for a typical STM image and Fig. S4 for more experimental data). Some of the LL peaks depend on the square-root of magnetic field, as shown in Fig. 4(b), the other LLs show linear field dependence, as shown in Fig. 4(c). However, the field dependence of the LLs corresponding to the massive Dirac fermions in both electron and hole sectors extrapolates to the same zero-field value [Fig. 4(c)], suggesting that, within experimental error, the two quartets, $LL_{(0,1,+)}$ and $LL_{(0,1,-)}$, are degenerate. This quite differs from that observed in the Bernal bilayer (Fig. 3), where a finite band gap is generated between the conduction and valence bands. For facility, the LL spectrum of the massive Dirac fermions in this case can be described by [3]

$$E_n = \text{sgn}(n)\hbar\omega_c\sqrt{|n|(|n|+1)} + E_0, \qquad n = ...-2,-1,0,1,2... \qquad (3)$$

It gives the same LL spectrum as that of Eq. (2) except the two degenerate spin-polarized quartets.

The result shown in Fig. 4(a-c) indicates that there are linear and parabolic subbands in the decoupled surface region, where the spectra are recorded, on HOPG. Similar STS spectra on HOPG were also reported by other groups [3,5], however, the origin of the coexistence of massless and massive Dirac fermions on the surface of graphite is still not very clear. Here, we attribute the unique spectra shown in Fig. 4(a) to the Landau quantization in Bernal graphene trilayer on graphite surface. In Bernal graphene trilayer, the low-energy band structure, in the simplest approximation (only the nearest-neighbor intra- and interlayer hopping parameters are taking into account), consists of single-layer-graphene-like and Bernal-bilayer-like subbands [23-27], as shown in Fig. 4(d) and Fig. 4(e). The expected Landau quantization in the Bernal trilayer is in excellent agreement with the experimental result. Theoretically, the charge neutrality points of the single-layer-graphene-like and Bernal-bilayer-like subbands in the Bernal trilayer are at the same energy in the simplest approximation. In the experiment, there is a slight energy difference, within 10 meV (see Supplementary Fig. S5 for the summarized result of nine different samples), between the $n = 0$ LL of massless and massive Dirac fermions. Such a subtle difference is very reasonable to be observed in the Bernal trilayer when the other non-nearest-neighbor hopping parameters (see Supplementary Fig. S6) play a role in affecting its band structure [23]. Here, we should point out that the observation of decoupled Bernal trilayer regions on graphite surface is also reasonable with taking into consideration of the fact that the main part of HOPG is the Bernal graphite. This, together with the quantitative agreement between our experiments and calculations provide indirect but compelling evidence that the decoupled Bernal trilayer accounts for the emergence of both massless and massive Dirac fermions on graphite surface.

Fitting the LL peaks to Eq. (1) and Eq. (3), as shown in Fig. 4(b) and Fig. 4(c), can obtain the Fermi velocity of massless Dirac fermions and the effective mass of massive Dirac fermions, respectively. Figure 5(a) summarizes the $v_F$ and $m^*$ obtained in nine different Bernal trilayers on graphite surface. Here, the sample indices are indexed from 1 to 9 with increasing the Fermi velocity, which is deduced from the Landau-level spectroscopy. It is interesting to note that the value of $v_F$ in different Bernal trilayers could differ more than 30% from each other. Similar large range of $v_F$ values was also observed in different graphene monolayer on graphite surface [4,9]. For example, it is measured to be $v_F = 0.79 \times 10^6$ m/s in ref. [4], to be $v_F = 1.10 \times 10^6$ m/s in the sample of ref. [9], and in our sample reported here the Fermi velocity of graphene monolayer could be as large as $v_F = 1.21 \times 10^6$ m/s.

Moreover, a strong correlation between $v_F$ and $m^*$ is observed: the value of $m^*$ decreases generally with the increasing of $v_F$. The result shown in Fig. 5(a) implies a high correlation between the massless and massive Dirac fermions on graphite surface. It is reasonable to assume that there is a common origin that affects the $v_F$ and $m^*$ in the Bernal trilayer simultaneously. To explore this behavior further, we calculate the band structure of the Bernal trilayer with different hopping parameters. The values of $v_F$ and $m^*$ depend sensitively on the hopping parameters. For example, the increase of the nearest-neighbor intralayer hopping parameter in the Bernal trilayer could simultaneously result in the opposite variations of the $v_F$ and $m^*$, i.e., the value of $v_F$ increases whereas the $m^*$ decreases, as shown in Fig. 5(b). This effect can quanlitatively explain our experimental result. Theoretically, the nearest-neighbor intralayer hopping parameter depends exponentially on the lattice constant. Although our experimental result really demonstrates a decreasing trend of the lattice constant from sample 1 to 8 (see Supplementary Fig. S7), it is unlikely that such a slight variation of the lattice constant, ~ 1%, could account for the observed behavior completely. Further studies would be required for a complete understanding of the behavior shown in Fig. 5(a).

In conclusion, we demonstrate that it is possible to identify decoupled graphene monolayer, Bernal bilayer, and Bernal trilayer on HOPG surface by their tunnelling spectra in high magnetic field. Therefore, the surface of graphite provides a natural ideal platform to probe the electronic spectra of graphene layers.

**References**

1. Zhou, S. Y., Gweon, G.-H., Graf, J., Fedorov, A. V., Spataru, C. D., Diehl, R. D., Kopelevich, Y., Lee, D.-H., Louie, S. G., Lanzara, A., First direct observation of Dirac fermions in graphite. *Nat. Phys.* **2**, 595-599 (2006).
2. Matsui, T., Kambara, H., Niimi, Y., Tagami, K., Tsukada, M., Fukuyama, H., STS observations of Landau levels at graphite surfaces. *Phys.Rev. Lett.* **94**, 226403 (2005).
3. Li, G., Andrei, E. Y., Observation of Landau levels of Dirac fermions in graphite. *Nat. Phys.* **3**,

623-627 (2007).

4. Li, G., Luican, A., Andrei, E. Y., Scanning tunneling spectroscopy of graphene on graphite. *Phys.Rev. Lett.* **102**, 176804 (2009).

5. Niimi, Y., Kambara, H., Fukuyama, H., Localized distributions of quasi-two-dimensional electronic states near defects artificially created at graphite surfaces in magnetic fields. *Phys.Rev. Lett.* **102**, 026803 (2009).

6. Miller, D. L., Kubista, K. D., Rutter, G. M., Ruan, M., de Heer, W. A., First, P. N., Stroscio, J. A., Observing the quantization of zero mass carriers in graphene. *Science* **324**, 924-927 (2009).

7. Orlita, M., Faugeras, C., Schneider, J. M., Martinez, G., Maude, D. K., Potemski, M., Graphite from the viewpoint of Landau level spectroscopy: an effective graphene bilayer and monolayer. *Phys.Rev. Lett.* **102**, 166401 (2009).

8. Li, G., Luican, A., Lopes dos Santos, J. M. B., Castro Neto, A. H., Reina, A., Kong, J., Andrei, E. Y., Observation of van Hove singularities in twisted graphene layers. *Nat. Phys.* **6**, 109-113 (2010).

9. Luican, A., Li, G., Reina, A., Kong, J., Nair, R. R., Novoselov, K. S., Geim, A. K., Andrei, E. Y., Single-layer behavior and its breakdown in twisted graphene bilayer. *Phys. Rev. Lett.* **106**, 126802 (2011).

10. Yin, L.-J., Qiao, J.-B., Wang, W.-X., Chu, Z.-D., Zhang, K. F., Dou, R.-F., Gao, C. L., Jia, J.-F., Nie, J.-C., He, L., Tuning structures and electronic spectra of graphene layers with tilt grain boundaries. *Phys. Rev. B* **89**, 205410 (2014).

11. Yin, L.-J., Qiao, J.-B., Xu, R., Dou, R.-F., Nie, J.-C., He, L., Electronic structures and their Landau quantizations in twisted graphene bilayer and trilayer. arXiv: 1410.1621.

12. Xu, R., Yin, L.-J., Qiao, J.-B., Bai, K.-K., Nie, J.-C., He, L., Dirac probing of the stacking order and electronic spectrum of rhombohedral trilayer graphene with scanning tunneling microscopy. arXiv: 1409.5933. to appear in *Phys. Rev. B*

13. Luk'yanchuk, I. A., Kopelevich, Y., Phase analysis of quantum oscillations in graphite. *Phys. Rev. Lett.* **93**, 166402 (2004).

14. Gusynin, V. P., Sharapov, S. G., Unconventional integer quantum Hall effect in graphene. *Phys. Rev. Lett.* **95**, 146801 (2005).

15. Yan, W., Liu, M., Dou, R.-F., Meng, L., Feng, L., Chu, Z.-D., Zhang, Y., Liu, Z., Nie, J.-C., He, L., Angle-dependent van Hove singulariteis in a slightly twisted graphene bilayer. *Phys. Rev. Lett.* **109**, 126801 (2012).

16. Yan, W., He, W.-Y., Chu, Z.-D., Liu, M., Meng, L., Dou, R.-F., Zhang, Y., Liu, Z., Nie, J.-C., He, L., Strain and curvature induced evolution of electronic band structures in twisted graphene bilayer. *Nature Commun.* **4**, 2159 (2013).

17. Rutter, G. M., Jung, S., Klimov, Klimov, N. N., Newell, D. B., Zhitenev, N. B., Stroscio, J. A., Mocroscopic polarization in bilayer graphene. *Nature Phys.* **7**, 649 (2011).

18. Song, Y. J., Otte, A. F., Kuk, Y., Hu, Y., Torrance, D. B., First, P. N., de Heer, W. A., Min, H., Adam, S., Stiles, M. D., MacDonald, A. H., Stroscio, J. A., High-resolution tunnelling spectroscopy of a graphene quartet. *Nature* **467**, 185 (2010).

19. McCann, E., Fal'ko, V. I., Landau-level degeneracy and quantum Hall effect in a graphite bilayer. *Phys. Rev. Lett.* **96**, 086805 (2006).

20. Zhang, Y., Tang, T.-T., Girit, C., Hao, Z., Martin, M. C., Zettl, A., Crommie, M. F., Shen, Y. R., Wang, F., Direct observation of a widely tunable bandgap in bilayer graphene. *Nature* **459**, 820 (2009).

21. Ohta, T., Bostwick, A., Seyller, T., Horn, K., Rotenberg, E., Controlling the electronic structure of bilayer graphene. *Science* **313**, 951 (2006).

22. Zhou, S. Y., Gweon, G.-H., Fedorov, A. V., First, P. N., de Heer, W. A., Lee, D.-H., Guinea, F., Castro Neto, A. H., Lanzara, A., Substrate-induced bandgap opening in epitaxial graphene. *Nature Mater.* **6**, 770 (2007).

23. Yuan, S., Roldan, R., Katsnelson, M. I., Landau level spectrum of ABA- and ABC-stacked trilayer graphene. *Phys. Rev. B* **84**, 125455 (2011).

24. Taychatannapat, T., Watanabe, K., Taniguchi, T., Jarillo-Herrero, P., Quantum Hall effect and

Landau-level crossing of Dirac fermions in trilayer graphene. *Nature Phys.* **7**, 621 (2011).

25. Bao, W., Jing, L., Jr, J. V., Lee, Y., Liu, G., Tran, D., Standley, B., Aykol, M., Cronin, S. B., Smirnov, D., Koshino, M., McCann, E., Bockrath, M., Lau, C. N., Stacking-dependent band gap and quantum transport in trilayer graphene. *Nature Phys.* **7**, 948 (2011).

26. Yankowitz, M., Wang, F., Lau, C. N., LeRoy, B. J., Local spectroscopy of the electrically tunable band gap in trilayer graphene. *Phys. Rev. B* **87**, 165102 (2013).

27. Yankowitz, M., Wang, J. I-J., Birdwell, A. G., Chen, Y.-A., Watanabe, K., Taniguchi, T., Jacquod, P., San-Jose, P., Jarillo-Herrero, P., LeRoy, B. J., Electric field control of soliton motion and stacking in trilayer graphene. *Nature Mater.* **13**, 786 (2014).

**Acknowledgments**

This work was supported by the National Basic Research Program of China (Grants Nos. 2014CB920903, 2013CBA01603, 2013CB921701), the National Natural Science Foundation of China (Grant Nos. 11422430, 11374035, 11474022, 51172029, 91121012), the program for New Century Excellent Talents in University of the Ministry of Education of China (Grant No. NCET-13-0054), Beijing Higher Education Young Elite Teacher Project (Grant No. YETP0238).

**Figure Legends:**

**Figure 1.** (a) and (b), Two typical atomic resolution STM images recorded on graphite surface. We can observe clear hexagonal lattice in panel (a) and triangular lattice in panel (b). (c), Four different *dI/dV-V* spectra recorded on different regions of graphite surface under the magnetic field of 8 T.

**Figure 2.** (a) STS spectra of decoupled graphene monolayer on graphite surface for various magnetic fields. LL indices are marked. For clarity, the curves are offset in Y-axis and all the spectra are shifted to make the n = 0 LL stay at the same bias. (b) LL peak energies for magnetic fields of 1 to 8 T versus square root of LL index and applied field, as expected for massless Dirac fermions. The solid line is a linear fit of the data with Eq. (1). Inset: the energy of the $n = 0$ LL at different magnetic fields. The solid line is a linear fit to the data points. Similar shift of the $n = 0$ LL with magnetic field is also observed previously in graphene multilayer and is attributed to the redistribution of charge in the multilayer [6,18]. (c) LL peak energies for different indices $n$ show the square-root dependence on magnetic field. The solid lines are the fits with Eq. (1).

**Figure 3.** (a) STS spectra of decoupled graphene Bernal bilayer on graphite surface for various magnetic fields. The band gap, ~ 35 meV, and LL indices are marked. The $LL_{(0,1,+)}$ and $LL_{(0,1,-)}$ projected on the top layer and the underlayer graphene, respectively. (b) LL peak energies plotted against $+(n(n-1))^{1/2}B$ for conduction band and $-(n(n-1))^{1/2}B$ for valence band. The red lines are the linear fit of the data with Eq. (2). The inset shows schematic structure of LLs in bilayer graphene in the quantum Hall regime with a finite bandgap. (c) LL peak energies for different indices $n$ show the linear dependence on magnetic field.

**Figure 4.** (a) A series of tunnelling spectra acquired on a graphite surface. The peaks marked with blue

arrows correspond to LLs of massless Dirac fermions. The peaks marked with red arrows correspond to LLs of massive Dirac fermion. The dotted line represents the charge neutral point. (b) LL peak energies of massless Dirac fermion shows a linear dependence against $\text{sgn}(n)(|n|/B)^{1/2}$. The solid line is a linear fit of the data with the slope yielding a Fermi velocity of $v_F = (0.986 \pm 0.006) \times 10^6$ m/s. Inset: LL peak energies for different indices $n$ versus the square-root of magnetic field. The solid lines are linear fits of the data. (c), LL peak energies of massive Dirac fermions shows a linear dependence against $\text{sgn}(n)(|n|/(|n|+1))^{1/2}B$. The solid line is a linear fit of Eq. (3) yielding the effective mass $m^* = (0.031 \pm 0.002)m_e$. Inset: LL peak energies for different indices $n$ show linear dependence on the magnetic field. According to our experimental result, there is an energy difference of about 9 meV between the $n = 0$ LL of the massless and massive Dirac fermions. (d) and (e), Schematic low-energy band structure of graphene Bernal trilayer around the $K$ point under zero magnetic field and in the quantum Hall region.

**Figure 5.** (a) The Fermi velocity of massless Dirac fermions and effective mass of massive Dirac fermions obtained in different Bernal trilayers. The open circle (sample 9) is the experimental data reported in ref. [3]. The sample indices are indexed from 1 to 9 with increasing the Fermi velocity. (b), Low-energy band structure of Bernal graphene trilayer for different nearest-neighbor intralayer hopping parameters (here, only the nearest-neighbor intra- and interlayer hopping parameters are considered).

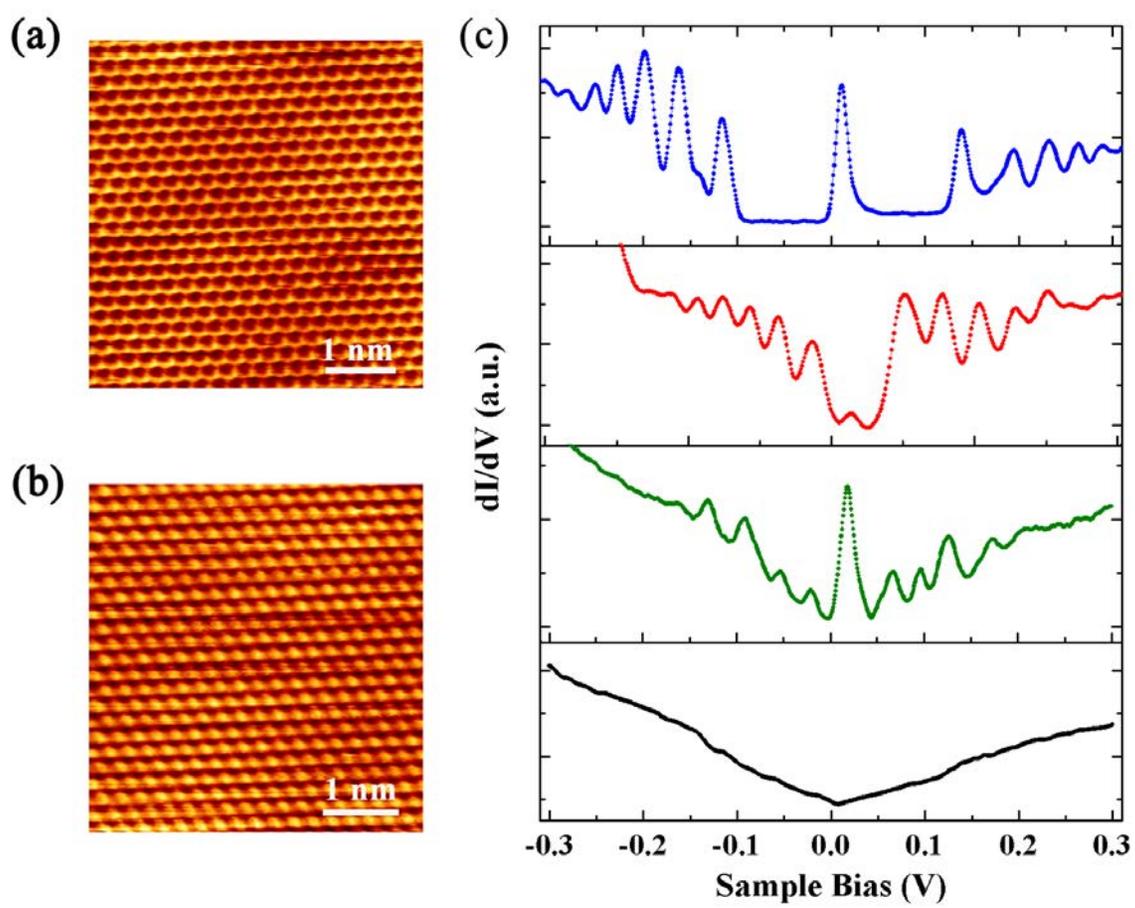

Figure 1

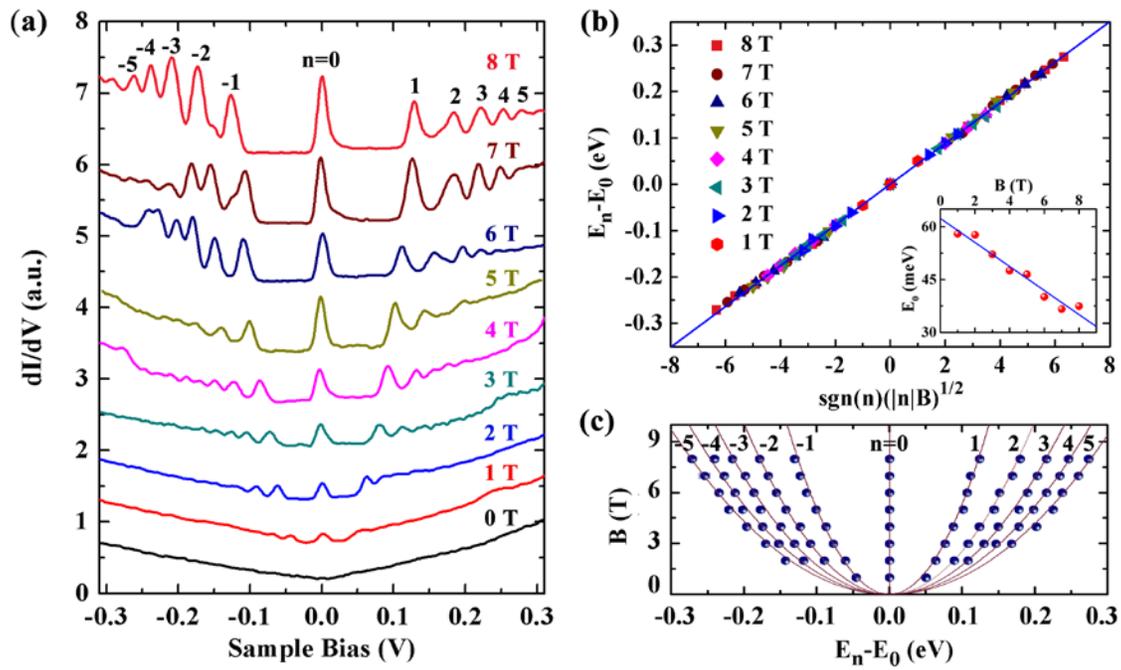

Figure 2

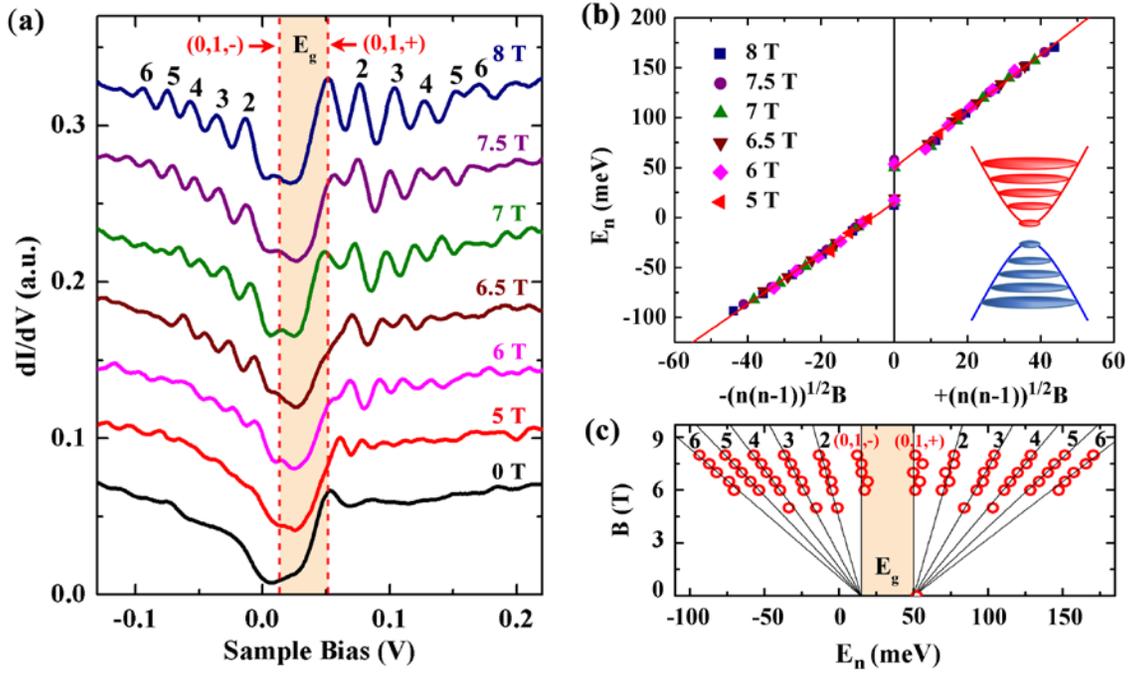

Figure 3

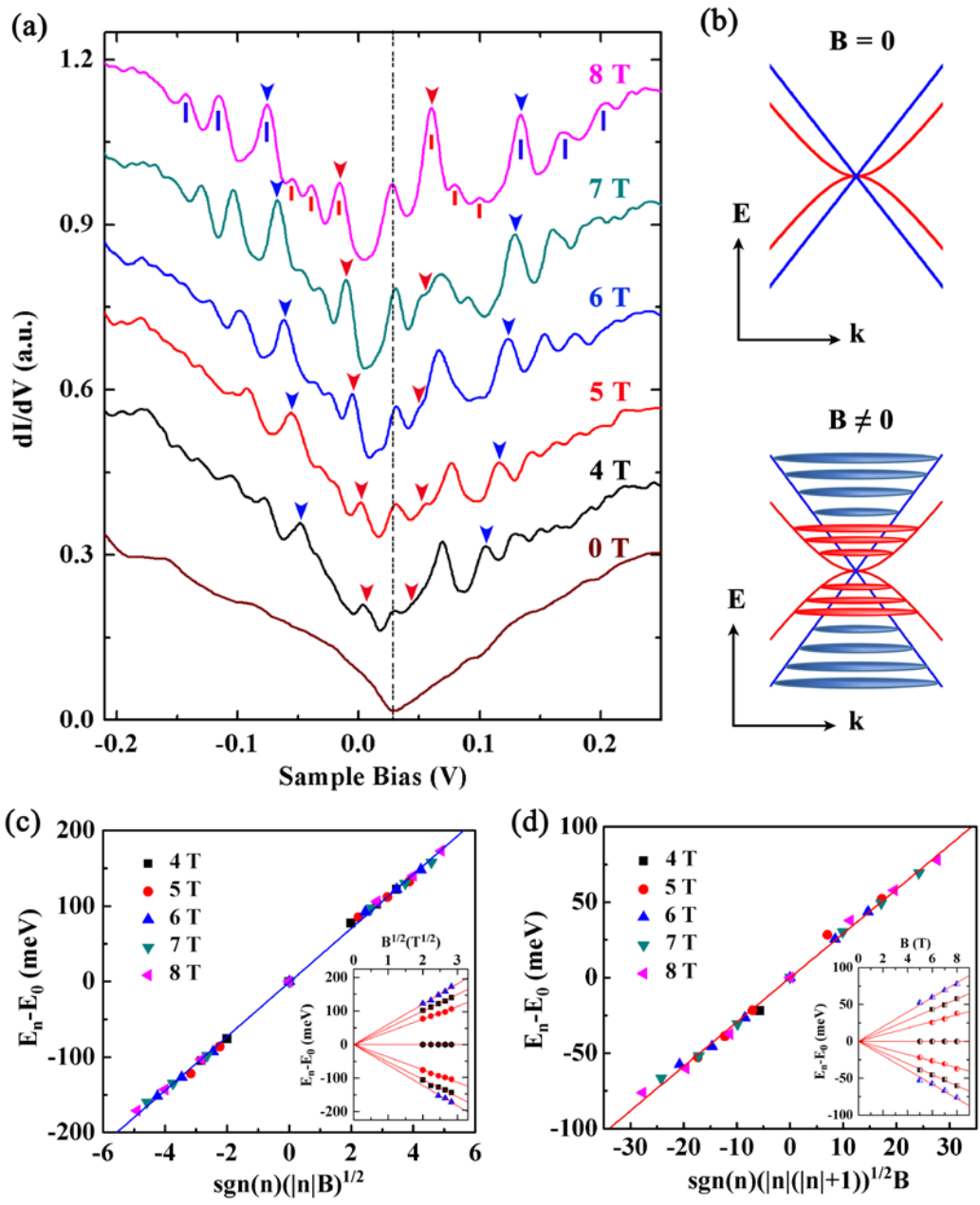

Figure 4

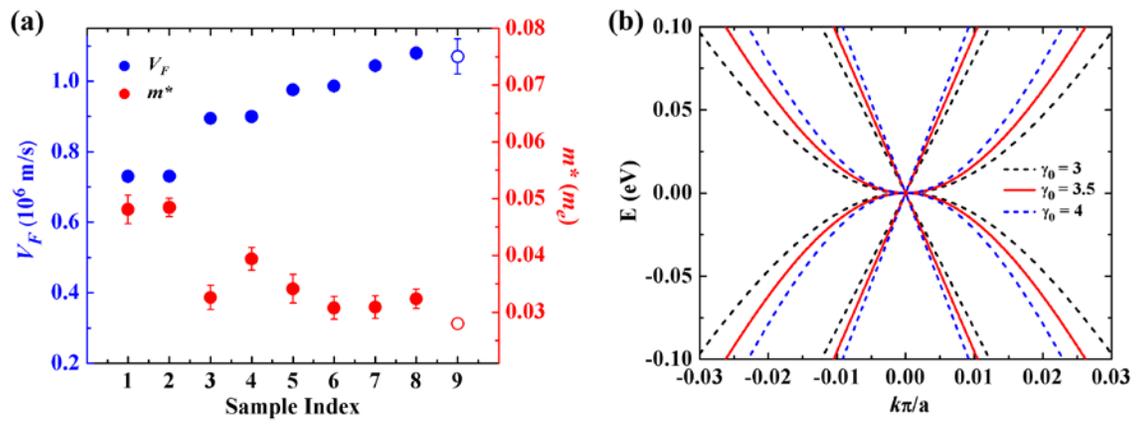

Figure 5